\documentclass[reprint,aps,prl,superscriptaddress,nofootinbib]{revtex4-2}

\usepackage{amsmath,amssymb,bm}
\usepackage{graphicx}
\usepackage{tikz}
\usetikzlibrary{arrows.meta,calc,decorations.pathmorphing}
\usepackage{xcolor}
\usepackage[normalem]{ulem}
\usepackage[colorlinks=true,linkcolor=blue,citecolor=blue,urlcolor=blue]{hyperref}

\newcommand{\ii}{\mathrm{i}}

\newcommand{\hc}{\mathrm{H.c.}}
\newcommand{\ket}[1]{|#1\rangle}
\newcommand{\bra}[1]{\langle #1|}

\newcommand{\sgn}{\operatorname{sgn}}

\begin{document}

\title{Synthetic Berry curvature in atom-cavity systems}

\author{Zheng Tang}
\thanks{These authors contributed equally to this work.}
\affiliation{State Key Laboratory of Quantum Optics Technologies and Devices, Institute of Theoretical Physics, Shanxi University, Taiyuan 030006, China}
\author{Rui-Lin Zhang}
\thanks{These authors contributed equally to this work.}
\affiliation{Hefei National Research Center for Physical Sciences at the Microscale and School of Physical Sciences, 
University of Science and Technology of China, Hefei 230026, China}
\author{Xiaotian Nie}
\affiliation{Intelligent Quantum Inception Co., Ltd., Haidian, Beijing, 100083, China}
\affiliation{iFLYTEK Research, Hefei, 230088, China}
\author{Li Chen}
\email{lchen@sxu.edu.cn}
\affiliation{State Key Laboratory of Quantum Optics Technologies and Devices, Institute of Theoretical Physics, Shanxi University, Taiyuan 030006, China}
\author{Wei Zheng}
\email{zw8796@ustc.edu.cn}
\affiliation{Hefei National Research Center for Physical Sciences at the Microscale and School of Physical Sciences, 
University of Science and Technology of China,
Hefei 230026, China}
\affiliation{CAS Center for Excellence in Quantum Information and Quantum Physics,
University of Science and Technology of China, Hefei 230026, China}
\affiliation{Hefei National Laboratory, 
University of Science and Technology of China, Hefei 230088, China}

\date{\today}

\begin{abstract}
In atom-cavity systems, mean field theory is widely used, in which the quantum cavity operator is replaced by a classical amplitude. Then the problem is converted into atoms moving in a self-consistent potential. The mean field treatment captures the physics of cavity mediated interactions, and predicts the self-organized superradiant phase. In this work, however, we show that it fails for certain atom-cavity coupling: it predicts zero ground state atomic current where the fully quantum calculation exhibits a finite one. We find that the origin of this failure is the non-zero Berry curvature in the synthetic dimension spanned by the photon Fock ladder and real space. In this synthetic picture, the atomic current can be understood as Hall response of cavity detuning, whereas mean field collapses this dimension and discards the atom--photon correlations required for its Hall response. The same beyond-mean-field geometry predicts Laughlin-like photon generation under adiabatic flux insertion. Our work identifies synthetic Berry curvature as a concrete mechanism for the breakdown of static mean field in atom--cavity physics.
\end{abstract}

\maketitle

\emph{Introduction.--}
When cold atoms are coupled to quantized cavity modes, the properties of both the atomic gas and the cavity field are substantially modified~\cite{Ritsch2002Cooling,ritsch2013cold,mivehvar2021cavity}. The cavity field can mediate infinite-range interactions, monitor the atomic order, and participate in the ordered state itself, enabling non-trivial steady states like Dicke superradiance~\cite{baumann2010dicke,WuHaiBing2021FermionSuperradiance,Nagy2010SR,Zhai2014SR-Fermi,Keeling2014SR-Fermi,Piazza2014SR-Fermi,Lv2024-SR,Zhang2021-SR-RabiTriangle,Zhang2022-SR-RabiRing}, self-organized density waves~\cite{Hemmerich2015SR-SF-MI,Esslinger2016MI-Cavity}, and supersolid-like phases~\cite{Esslinger2012Roton,Lev2010SuperSolid}, as well as new non-steady phases, such as continuous time crystals~\cite{Hemmerich2022CTC,Keeling2010LC,Risch2015LC} and self-organized topological pumping~\cite{dreon2021self}. Atom--cavity systems therefore offer a route to quantum simulators in which geometry, dissipation, and interaction are engineered via dynamical photons rather than imposed as static potentials.

A common theoretical starting point is the semiclassical or mean-field approximation, in which the cavity mode is assumed to be in the coherent state, such that the photon operator can be replaced by its mean value, a $c$-number, $\hat a(t) \to \alpha(t) = \langle \hat{a}(t) \rangle$. This converts the atom--cavity problem into atoms moving in a self-consistent optical potential. The mean cavity field $\alpha(t)$ is determined by the feedback of atoms. Such mean field treatment successfully captures the physics of cavity mediated infinite-range interactions, and can predict the self-organized states like superradiance, and dynamical phases such as time crystals.

\begin{figure*}[t]
\centering
\includegraphics[width=\textwidth]{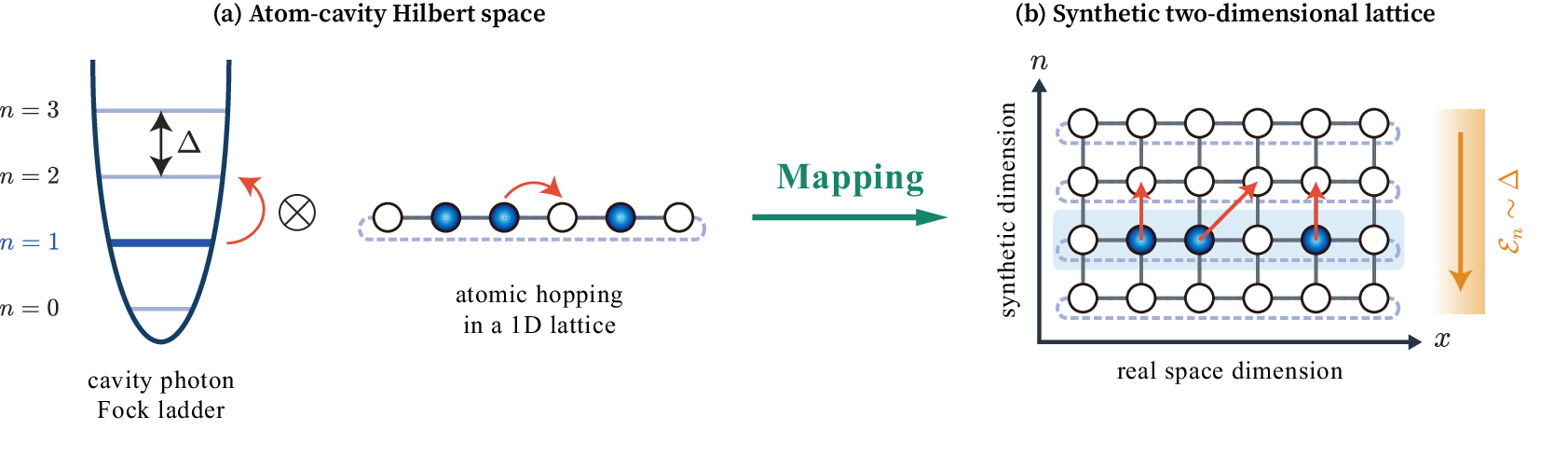}
\caption{Mapping from an atom--cavity system to a synthetic two-dimensional lattice. (a) The physical Hilbert space is the direct product of the cavity photon Fock ladder and many-atom Hilbert space in one-dimensional lattice. The detuning term $\Delta\hat a^\dagger\hat a$ separates neighboring photon sectors, while cavity-assisted hopping changes the atomic position together with the photon state. (b) In the single-particle representation the basis $|i,n\rangle$ forms a synthetic lattice whose horizontal direction is real space and whose vertical direction is the photon-number coordinate. The detuning acts as a uniform electric field $\mathcal{E}_n \varpropto \Delta$ along the synthetic direction, and the atomic current is interpreted as a Hall response generated by Berry curvature in this synthetic plane.}
\label{fig:synthetic_mapping}
\end{figure*}

However, does the mean field approximation remain valid for all types of atom-cavity couplings? In this Letter, we reveal a regime where this widely trusted approximation leads to a qualitatively incorrect ground state. Specifically, we find that for a certain cavity-assisted hopping and staggering configuration, the fully quantum solution supports a finite equilibrium atomic current, while the mean field theory inevitably predicts a vanishing current. This is not merely a quantitative correction—it signals a fundamental breakdown of the mean field picture.  The resolution, as we will show, lies in the Berry curvature in the synthetic dimension lattice generated by correlation between atomic motion in real space and photon Fock states. Therefore the ground state atomic current could be understood as the Hall response to the cavity detuning (See Fig.~\ref{fig:synthetic_mapping}). Remarkably, even though there is no interaction between atoms in real space, their motions in synthetic lattice are not independent. Since atoms in the synthetic lattice share the same photon coordinate, that imposes a strong constraint on the hopping of atoms.

The mean field approximation, however, by collapsing the entire photon ladder into a single quadrature point, erases this geometric structure and thus misses the associated Hall-type response. Therefore, the geometry of the photon coordinate is essential for understanding the response of the system. Based on this synthetic-dimension mapping, we predict a Laughlin-type photon generation mechanism~\cite{laughlin1981quantized}. As a flux is adiabatically inserted through the real-space ring, the Hall response along the synthetic direction will generate photons even after the Hamiltonian returns to itself.

\emph{Model.--}
We consider spinless fermions on a bipartite chain of $L$ unit cells, coupled to one quantized cavity mode. The sublattice is denoted by $\tau=A,B$, and $i + L \equiv i$ implements periodic real-space boundary condition. The Hamiltonian is given by
\begin{align}
\hat H  &=   \Delta \hat a^\dagger \hat a  + \bigl(m+\delta m\hat{P}\bigr)\hat{D} \nonumber\\
         & - \bigl(h-\delta h\hat{Q}\bigr)\hat{K}_{0} - \bigl(h+\delta h\hat{Q}\bigr)\hat{K}_{1}, 
\label{eq:model}
\end{align}
where $\hat c_{i\tau}$ annihilates an atom in cell $i$ and $\hat a$ annihilates a cavity photon. Here $\hat Q=(\hat a+\hat a^\dagger)/2$ and $\hat P=(\hat a-\hat a^\dagger)/2i$ are two cavity quadratures, while $\Delta$ is the cavity detuning. The staggered density is $\hat{D}=\sum_i(\hat c_{iA}^\dagger\hat c_{iA}-\hat c_{iB}^\dagger\hat c_{iB})$; $\hat K_0=\sum_i(\hat c_{iB}^\dagger\hat c_{iA}+\hc)$ and $\hat K_1=\sum_i(\hat c_{i+1,A}^\dagger\hat c_{iB}+\hc)$ are the intracell and intercell hopping operators. We neglect cavity loss and focus on the ground state properties. The common mean field theory replaces the cavity operator by its mean value, $\hat a\to\alpha=\langle\hat a\rangle$, such that $\hat{Q} \to \mathrm{Re}\alpha$ and $\hat{P} \to \mathrm{Im}\alpha$. In this picture, atoms move in a Rice--Mele chain with real hopping amplitudes~\cite{rice1982elementary}, $\hat H_{\mathrm{RM}} (\alpha)  =  \bigl(m+\delta m \mathrm{Im} \alpha\bigr)\hat{D} - \bigl(h-\delta h\mathrm{Re} \alpha\bigr)\hat{K}_{0} - \bigl(h+\delta h\mathrm{Re} \alpha\bigr)\hat{K}_{1}.$
Note that the full quantum Hamiltonian (\ref{eq:model}) breaks time-reversal ($\hat{T} \hat{P} \hat{T}^{-1}=-\hat{P}$), so a finite equilibrium current is symmetry-allowed. However, after mean field replacement, $\hat{H}_{\rm RM}(\alpha)$ has purely real matrix elements, effectively restoring time-reversal symmetry and forcing zero current.

\begin{figure}[t]
\centering
\includegraphics[width=0.5\columnwidth]{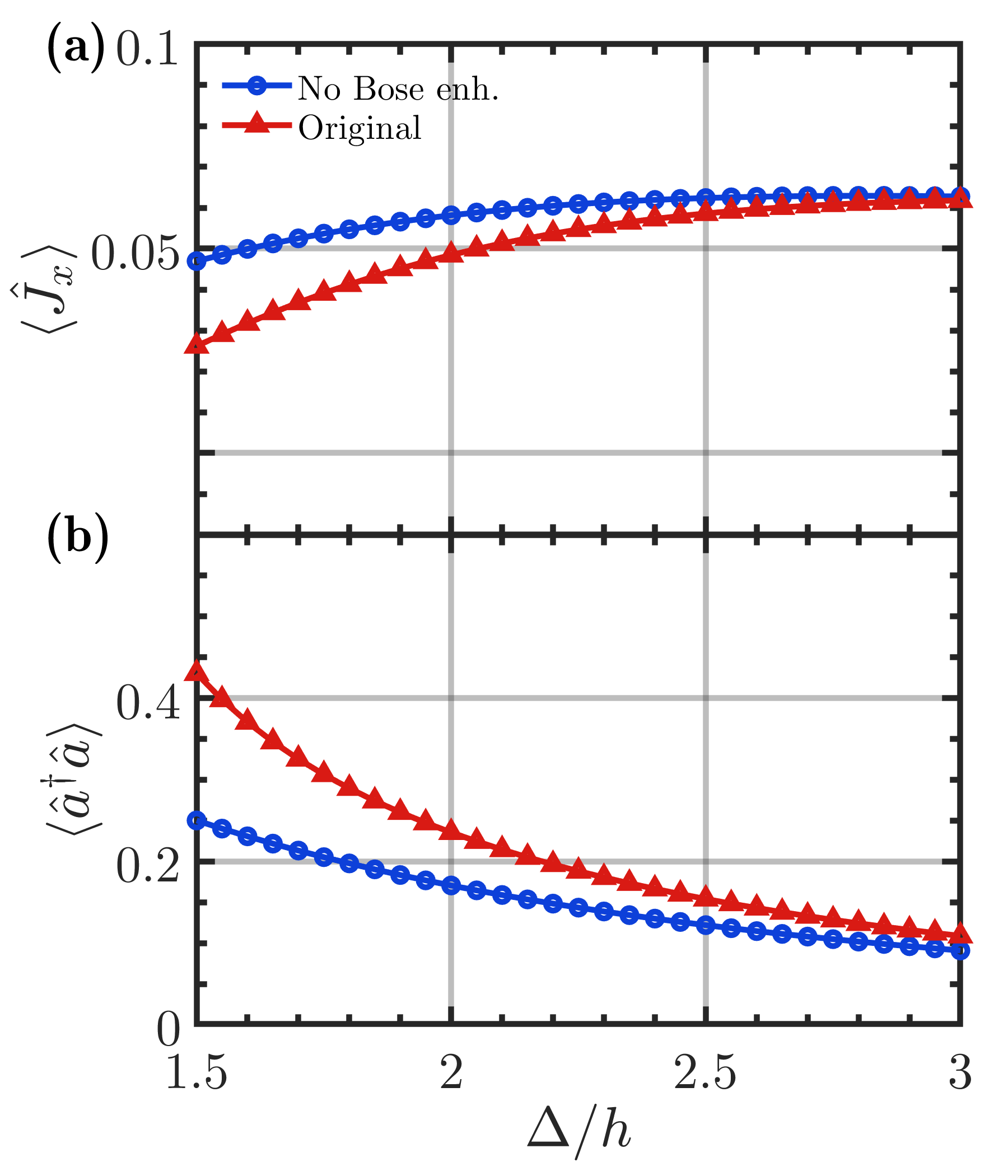}\hfill
\includegraphics[width=0.5\columnwidth]{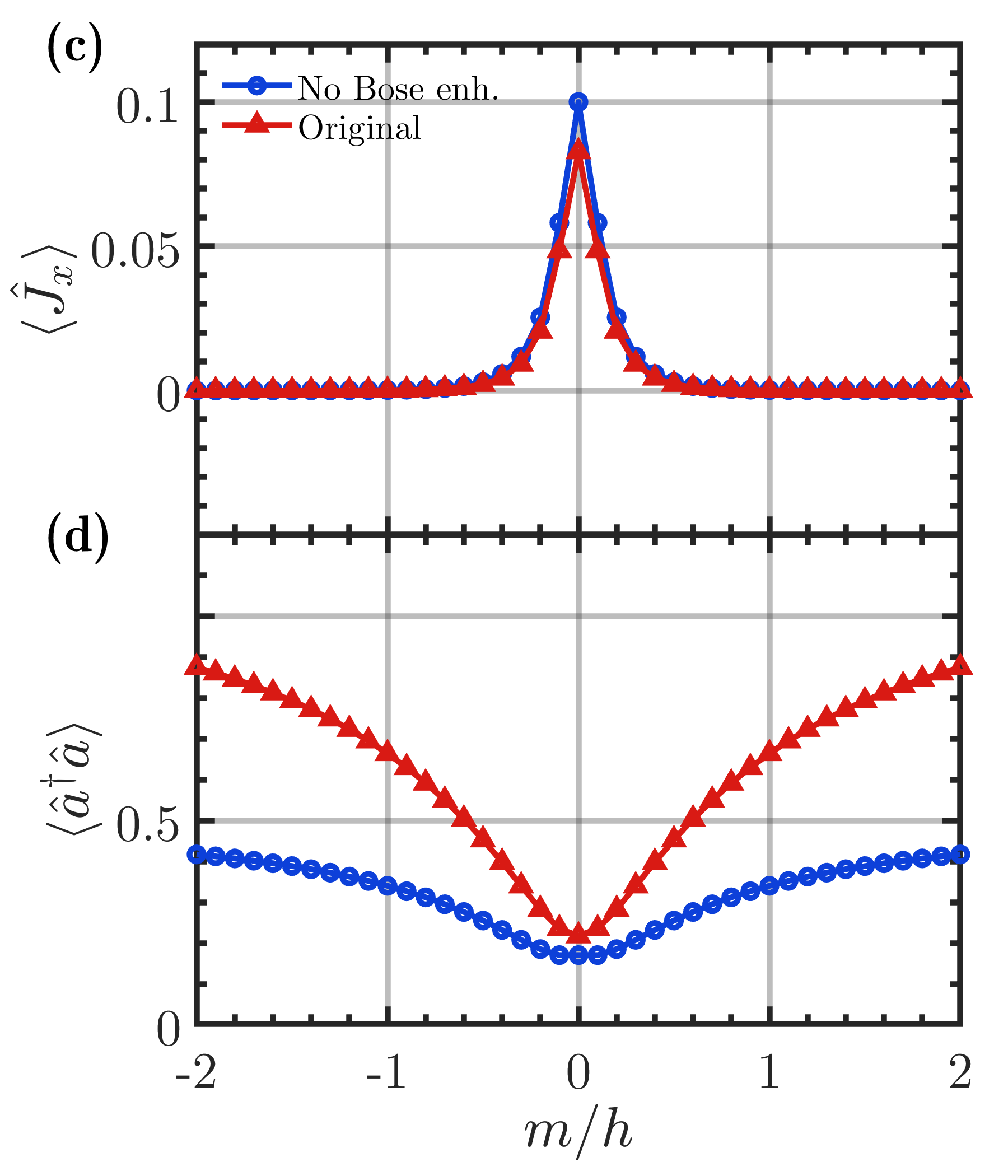}
\caption{Ground-state observables via exact diagonalization. (a,b) Atomic current and mean photon number versus cavity detuning at $m/h=0.1$. (c,d) The corresponding quantities versus $m/h$ at $\Delta/h=2$. The sharp current feature at $m/h=0$ follows from the jump of the selected collective synthetic momentum $q_*$ and its slice Hall conductance. The blue data shows the model with the no-Bose-enhancement. Here $L=6$, $\nu=0.5$, $\delta h /h =0.4$ and $\delta m/h = 0.7$.}
\label{fig:current}
\end{figure}

Exact diagonalization, however, yields a finite ground-state current $\langle \hat{J}_{x} \rangle \ne 0$ (see Fig.~\ref{fig:current}). Here the atomic current operator reads $\hat{J}_{x} = \hat{J}_{0} +\hat{J}_{1}$~\cite{SM}, where $\hat J_{0}   = \frac{\ii}{L}\sum_{i}( h - \delta h\hat Q )( \hat c_{iB}^\dagger  \hat c_{iA}  - h.c. )$ and $\hat{J}_{1} = \frac{\ii}{L}\sum_{i}( h + \delta h\hat Q )( \hat c_{i + 1A}^\dagger  \hat c_{iB}  - h.c.)$. The current grows with detuning even as the photon number decreases, and  develops a sharp feature at $m=0$. Its persistence after removing Bose enhancement effect ($\hat{a}=\sum_{n=0}^{\infty}\sqrt{n}\ket{n-1}\bra{n} \rightarrow \sum_{n=0}^{\infty}\ket{n-1}\bra{n}$), identifies atom--photon correlations, rather than large photon occupation, as the relevant mechanism. These observations pose a sharp question: how can a time-independent Hamiltonian support an equilibrium current although the corresponding mean-field treatment fails? Below we show that the answer lies in the geometry in the synthetic dimension.

\emph{Map to synthetic higher dimensional lattice.} To expose the synthetic coordinate, we map an $N$-fermion state in photon sector $n$ as
\begin{equation}
\frac{(\hat{a}^{\dagger})^{n}}{\sqrt{n!}}
\hat c_{x_1}^\dagger\cdots
\hat c_{x_N}^\dagger\ket{\mathrm{vac}}
\rightarrow
\hat{\tilde c}_{n,x_1}^\dagger
\cdots
\hat{\tilde c}_{n,x_N}^\dagger\ket{\mathrm{vac}}.
\label{eq:Mapping}
\end{equation}
Here $\ket{\mathrm{vac}}$ is the vacuum state; $x = (i,\tau)$ and $\hat{\tilde c}_{n,x}^\dagger$ creates a composite fermion carrying both the atomic orbital $x$ and the photon coordinate $n$.  The atomic Hamiltonian moves these composite fermions along the real-space direction, whereas the $\hat{a}$ and $\hat{a}^{\dagger}$ terms change the photon coordinate. Therefore one obtains the Hamiltonian in the synthetic two dimensional lattice as~\cite{SM}
\begin{align}
 \hat H_{\mathrm{syn}} &= \sum_{n = 0}^\infty   \sum_{i} \{ (\Delta n + m)\hat {\tilde{c}}_{n;iA}^\dagger  \hat{\tilde{c}}_{n;iA} \nonumber\\ 
 & + (\Delta n - m)\hat{\tilde{c}}_{n;iB}^\dagger  \hat{\tilde{c}}_{n;iB}  \nonumber\\ 
  &- h(\hat{\tilde{c}}_{n;iB}^\dagger  \hat{\tilde{c}}_{n;iA}  + \hat{\tilde{c}}_{n;i + 1A}^\dagger  \hat{\tilde{c}}_{n;iB}  + \hc) \nonumber\\ 
  &- \frac{{\ii\delta m}}{2}\sqrt n (\hat{\tilde{c}}_{n - 1;iA}^\dagger  \tilde c_{n;iA}  - \hat{\tilde{c}}_{n - 1;iB}^\dagger  \hat{\tilde{c}}_{n;iB}  - \hc) \nonumber\\ 
  &+ \frac{{\delta h}}{2}\sqrt n ( \hat{\tilde{c}}_{n - 1;iB}^\dagger  \hat{\tilde{c}}_{n;iA}  + \hat{\tilde{c}}_{n;iB}^\dagger  \hat{\tilde{c}}_{n - 1;iA}  + \hc \nonumber\\ 
  &- \hat{\tilde{c}}_{n - 1;i + 1A}^\dagger  \hat{\tilde{c}}_{n;iB}  - \hat{\tilde{c}}_{n;i + 1A}^\dagger  \hat{\tilde{c}}_{n - 1;iB}  + \hc) \}. 
\label{eq:synthetic_Ham}
\end{align}
Note that the cavity detuning $\Delta$ serves as linear potential in synthetic direction. The cavity-assisted hopping acts as the next-nearest neighbor hopping in the diagonal direction. The factor $\sqrt{n}$ accounts for the Bose-enhancement effect. 

Unlike the previously studied synthetic dimension in internal degree of freedom or frequency domain~\cite{Mancini2015synthetic,Spielman2015synthetic,Boada2012synthetic,Zilberberg2013synthetic,celi2014synthetic,Goldman2015synthetic,Lustig2019photonic,Dutt2020photonic,Fan2016photonic,Fan2018photonic,Ozawa2019Rev}, the photon coordinate here is collective: all atoms share the same Fock state.  Therefore the physical Hilbert space is not the full Fock space of independent $\hat{\tilde c}_{n,x}$ fermions, but the constrained subspace $\mathcal H_{\rm phys}={\rm span}
\left\{
\hat{\tilde c}_{n,x_1}^\dagger\cdots
\hat{\tilde c}_{n,x_N}^\dagger\ket{\mathrm{vac}}
\right\}$. This imposes the physical constraint that composite fermions
cannot occupy different photon coordinates simultaneously,
\begin{equation}
\hat{\tilde{c}}^{\dagger}_{n,i\tau}\hat{\tilde{c}}^{\dagger}_{n',i'\tau'}=0,
\qquad n \ne n',
\label{eq:constraint}
\end{equation}
This constraint is the main difference between the single-particle problem. Consequently a synthetic-direction hopping event must transfer the whole many-body configuration from one photon layer to another. 

\emph{Berry curvature in the synthetic lattice.}
To isolate the geometry in the synthetic lattice, we first neglect the Bose enhancement factor, $\sqrt{n} \to 1$, set $\Delta=0$ in Hamiltonian (\ref{eq:synthetic_Ham}), and impose periodic boundaries in both photon and real direction. These auxiliary assumptions restore translation symmetry along the synthetic direction; they are used only to expose the underlying band geometry in the momentum space. 

\begin{figure}[t]
\includegraphics[width=0.5\textwidth]{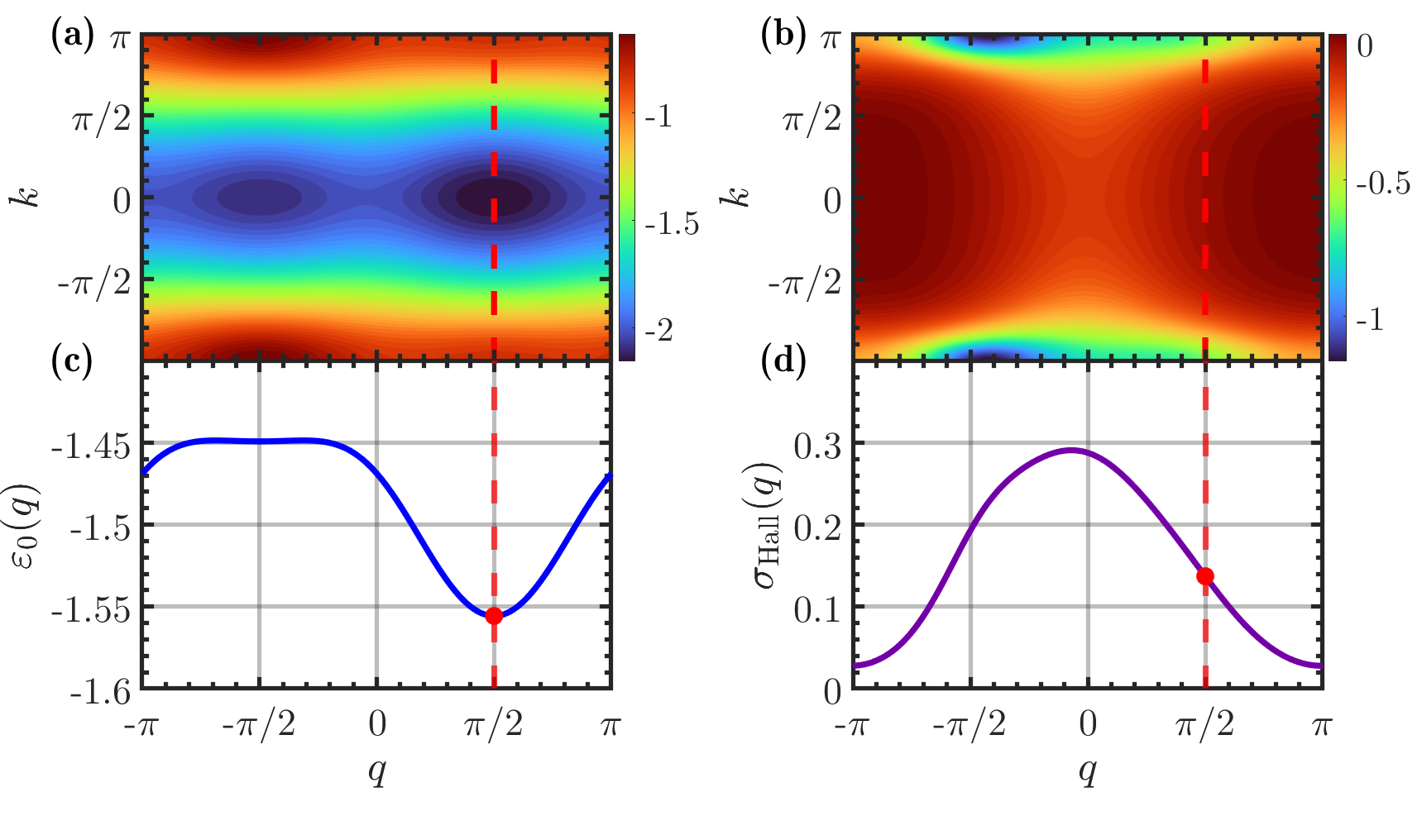}
\caption{Synthetic band and many-body block structure without Bose enhancement. (a) Single-atom lower-band energy $E_-(q,k)$, and (b) its Berry curvature. (c) Ground state energy in the $q$ block. The global ground state selects the minimum-energy block $q_*$, which is indicated by the dashed line. (d) Fixed-$q$ slice Hall conductance obtained from occupied-state Berry curvature. It is not quantized. Here $\nu=0.5$, $m/h =0.1$, $\delta h /h =0.4$ and $\delta m/h = 0.7$.}
\label{fig:SyntheticLattice_Band_Berry}
\end{figure}

As we argued, all atoms share the synthetic coordinate $n$, and hop collectively in the photon direction. Therefore, in the momentum space, all atoms populate one collective synthetic momentum $q$, conjugate to the collective synthetic coordinate $n$. That imposes the physical constraint for the composite fermions in momentum space, 
\begin{equation}
\hat{\tilde{c}}^{\dagger}_{q,k\tau}\hat{\tilde{c}}^{\dagger}_{q',k'\tau'} = 0,
\qquad q \ne q'. 
\label{eq:Momentumconstraint}
\end{equation}
As a result, the Hamiltonian in momentum space decomposes into blocks labelled by one collective synthetic momentum $q$,
\begin{equation}
\hat{H}_{\mathrm{syn}}=\bigoplus_q \hat H(q),
\qquad
\hat H(q)=\sum_k \hat\Psi_{q,k}^\dagger \mathcal H(q,k)\hat\Psi_{q,k}.
\label{eq:q_blocks}
\end{equation}
where $\hat\Psi_{q,k}=(\hat{\tilde{c}}_{q,kA},\hat{\tilde{c}}_{q,kB})^T$, and $\mathcal H(q,k) = \bm{d}(q,k) \cdot \bm{\sigma}$. The vector $\bm{d}(q,k)$ is given by
\begin{align}
d_x&=h-\delta h\cos q+(h+\delta h\cos q)\cos k,
\nonumber\\
d_y&=(h+\delta h\cos q)\sin k,
\nonumber\\
d_z& = m+\delta m \sin q.
\label{eq:d_vector}
\end{align}
The Hamiltonian in $q$ block $\hat{H}(q)$ can be diagonalized into  $\hat{H}(q) = \sum_{k\lambda} E_{\lambda}(q,k)\hat{\tilde{\gamma}}^{\dagger}_{q,k\lambda}\hat{\tilde{\gamma}}_{q,k\lambda}$, where
$E_\pm(q,k)=\pm |\bm d(q,k)|$ is the single-atom band energy. One can also obtain non-vanishing Berry curvature~\cite{xiao2010berry}, 
\begin{equation}
\Omega_{\pm}(q,k)
=\pm \bm d\cdot(\partial_k\bm d\times\partial_q\bm d)/2|\bm d|^3,
\label{eq:berry_curvature}
\end{equation}
The curve $(\delta h\cos q, m+\delta m\sin q)$ winds around the gap-closing point when $|m|<|\delta m|$. In that regime both the upper and the lower band are topological non-trivial, with Chern number $C_{\pm} = \pm \sgn(\delta h\delta m)$ in the convention of Eq.~\eqref{eq:berry_curvature}; otherwise $C_{\pm}=0$. Figure~\ref{fig:SyntheticLattice_Band_Berry}(a,b) displays the resulting lower-band dispersion and Berry-curvature texture.

\begin{figure}[t]
\centering
\includegraphics[width=0.5\textwidth,trim=0 12pt 0 4pt,clip]{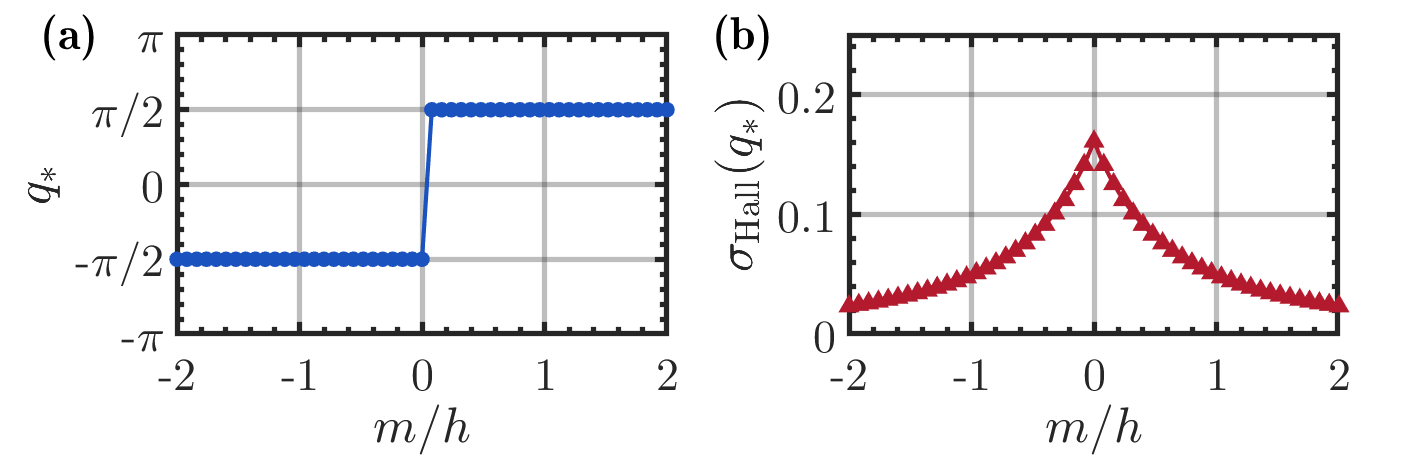}
\vspace{-0.5em}
\includegraphics[width=0.52\textwidth,trim=0 4pt 0 12pt,clip]{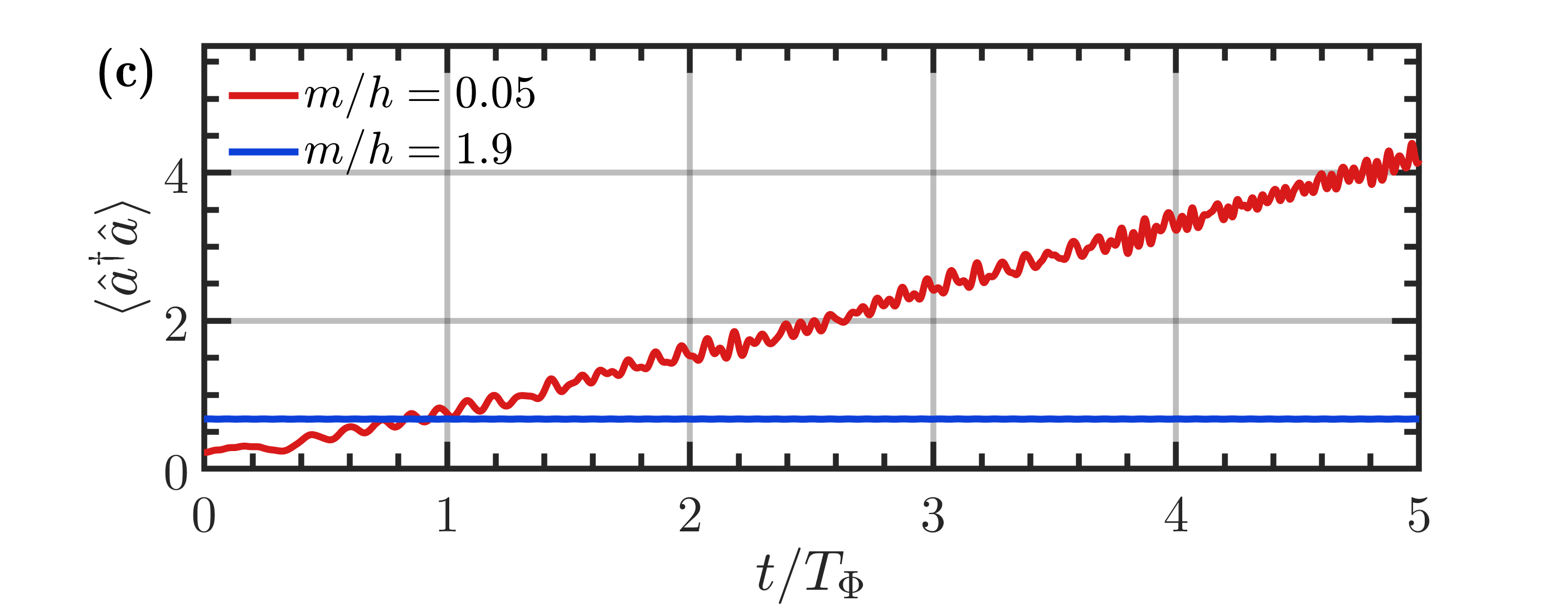}
\caption{(a) Ground-state collective synthetic momentum $q_*$ as a function of $m/h$, obtained by minimizing the half-filled many-body energy over the $q$ blocks. The selected block changes abruptly at $m/h=0$. (b) Hall conductance $\sigma_{\mathrm{Hall}}(q_*)$ evaluated in the ground state block. The jump of $q_*$ produces a cusp-like feature of $\sigma_{\mathrm{Hall}}(q_*)$ near $m=0$. (c) Photon number during a Laughlin-pumping protocol, where a flux is adiabatically threaded through the real-space ring. Here $L=4$, $\nu=0.5$, $\delta h /h =0.4$,  $\delta m/h = 0.7$, $\Delta/h = 1.5$, and pumping period is $T_{\Phi} = 20\pi$. }
\label{fig:SyntheticLattice_Band_Berry_MSweep}
\end{figure}

The many-body consequence differs crucially from an ordinary two-dimensional Fermi gas. All composite fermions must fill a common $q$ block, rather than independently occupying the energetically optimal $q$ at each $k$. The many-body ground state energy density in block $q$ is $\varepsilon_0(q)=\frac{1}{L}\sum_{k,\lambda=\pm} f_{\lambda,k}(q)E_{\lambda}(q,k)$, where $f_{\lambda,k}$ is the fermi distribution. At zero temperature the composite fermions fill the lowest $2L\nu$ single-particle levels. The ground-state block is selected by
\begin{equation}
q_*=\arg\min_q\varepsilon_0(q).
\label{eq:q_star}
\end{equation}
The corresponding slice Hall response is
\begin{equation}
\sigma_{\mathrm{Hall}}(q_*)
=-\sum_{k,\lambda=\pm}
f_{\lambda,k}(q_*)\Omega_{\lambda}(q_*, k),
\label{eq:slice_hall}
\end{equation}
At half filling, $f_{-,k}=1$ and $f_{+,k}=0$, hence Eq.~\eqref{eq:slice_hall} reduces to the Berry-curvature integral over a fixed-$q$ slice. It is generally not quantized, since one does not integrate over the full synthetic Brillouin zone~\cite{thouless1982quantized}. 
Therefore when $\Delta\ne 0$, a finite electric field $\mathcal{E}_{n}=\Delta/N$ is applied along the synthetic direction, produces a Hall current in the real direction~\cite{Halperin2017photonic},
\begin{equation}
\langle \hat{J}_{x} \rangle =\sigma_{\mathrm{Hall}}(q_{*})\mathcal{E}_{n}.
\label{eq:anomalous_velocity}
\end{equation}
Figure~\ref{fig:SyntheticLattice_Band_Berry_MSweep}(a,b) shows their mass dependence. At $m=0$, $q_*$ switches discontinuously, yielding a transition and a sharp feature in $\sigma_{\mathrm{Hall}}(q_*)$. It should be noted that this transition is not a topological transition, but a many-body block-selection level crossing. This explains the sharp structure of the exact-diagonalization current in Fig.~\ref{fig:current}(c). The block selection is a direct consequence of the shared photon coordinate and has no analogue for independently occupied synthetic momenta.

\emph{Why mean field misses the current.--}
Mean field freezes the cavity into a point $(\mathrm{Re}\alpha,\mathrm{Im}\alpha)$ in quadrature space and thereby collapses the photon direction and removes the Berry curvature encoded in Eq.~\eqref{eq:berry_curvature}. It retains the feedback of an average cavity field but discards correlations between atomic motion and different photon sectors. The failure can also be seen perturbatively. In the large-detuning limit, projecting onto the photon vacuum gives the effective Hamiltonian of atoms~\cite{SM}, $\hat{H}_{\mathrm{eff}} = (m\hat{D} - h\hat{K}_{+}) + \delta\hat{H}^{(2)}_{\mathrm{eff}}$, where 
\begin{align}
\delta\hat H_{\rm eff}^{(2)} &=
\frac{\ii\delta h\delta m}{4\Delta}\left[\hat{D},\hat{K}_{-} \right]-\frac{\delta h^2}{4\Delta}\hat{K}_{-}^2 -\frac{\delta m^2}{4\Delta}\hat{D}^2,
\label{eq:commutator_eff}
\end{align}
and $\hat K_\pm=\hat K_0\pm\hat K_1$. Note that $\ii\left[\hat{D},\hat{K}_{-} \right] \propto -\hat{\Pi} \hat{J}_{x} \hat{\Pi}$. Here $\hat{\Pi} = \ket{0}\bra{0} \otimes \hat{I}_{\mathrm{at}}$, is the projection operator to the zero-photon sector~\cite{SM}. Thus virtual photon processes generate a term directly proportional to the atomic current. That indicate a current carrying state will lower the energy of system. In the mean field treatment, the quadratures commute as numbers, hence its chiral contributions are absent.

\emph{Laughlin generation of photons.--}
The synthetic Hall picture predicts a dynamical consequence. Threading a slowly varying flux through the real-space ring generates an electric field along the ring that is proportional to increasing rate of the flux. It can be realized by impose a Peierls phase on the hopping term as $\hat K_0 \to\sum_i\big[e^{i\Phi(t)/2L}\hat c_{iB}^\dagger\hat c_{iA}+ \hc \big]$, and $\hat K_1 \to\sum_i\big[e^{i\Phi(t)/2L}\hat c_{i+1,A}^\dagger\hat c_{iB}+ \hc\big]$
where $\Phi(t) = 2\pi t/T_{\Phi} $, and $T_{\Phi}$ is the pumping period for one flux quanta. Due to the finite Hall conductance in the synthetic dimension, that will drive a current in the synthetic direction, $\hat{J}_{n} = \delta m \hat{Q} \hat{D} - \delta h \hat{P}\hat{K}_{-}$. That leads the increasing of the photon number~\cite{SM}, 
\begin{align}
  \frac{\mathrm{d} \langle \hat{n}_{\mathrm{ph}} \rangle}{{\mathrm{d}}t} =  \langle \hat{J}_n \rangle  =\sigma _{\mathrm{Hall}} \left( {q_{*},\Phi} \right)\frac{{{\mathrm{d}}\Phi }}{{\mathrm{d}}t},
\label{eq:photon_laughlin}
\end{align}
For a complete cycle, $\Delta \langle \hat{n}_{\mathrm{ph}} \rangle =\int_0^{2\pi}\mathrm d\Phi\,\sigma_{\mathrm{Hall}}(q_*,\Phi)$. Note that $\Delta \langle \hat{n}_{\mathrm{ph}} \rangle$ is generally not quantized, since the synthetic Hall conductance is not quantized. Flux insertion can therefore generate photons even though the Hamiltonian returns to itself: a cavity realization of Laughlin pumping with photon number as the transverse coordinate. Starting from the ground state at $\Phi(0) = 0$, we numerically calculate exact pumping dynamics, see Fig.~\ref{fig:SyntheticLattice_Band_Berry_MSweep}(c).  It shows the Laughlin photon generation at small $m$, and its suppression in the large $m$ regime.

\emph{Discussion.--}
The atom--cavity coupling in Eq.~(1) is not merely a formal model. A closely related coupling can be engineered with atoms inside an optical cavity driven by imbalanced pumps~\cite{Esslinger2021imbalanced,dreon2021self,ZW2023imbalanced}. In such a setup, the two pump pathways generate cavity-assisted hopping and density terms with independently tunable amplitudes and phases. when take in account the cavity loss, the current will survive in the steady state. However, the mechanism is beyond the synthetic Hall linear response picture of this work, which requires further investigation.

More broadly, this mean-field failure may be a general geometric mechanism, not a cavity-specific peculiarity. Analogous situations may arise wherever orbital, spin, gauge, or slave-particle sectors serve as internal spaces. Synthetic-dimension formulations thus offer a way to identify whether quantum fluctuations of an internal degree of freedom produce responses beyond classical self-consistent field.

\emph{Acknowledgment.--}
W.Z. acknowledges support from the NSF of China (Grants
No. GG2030007011 and No. GG2030040453) and Innovation
Program for Quantum Science and Technology (Grant No.
2021ZD0302004). L.C. acknowledges supports from the NSF of China (Grants No. 12574296, and No. U25A20198), the Sanjin Talent Program of Shanxi Province, and the fund for the Shanxi 1331 Project.


\begin{thebibliography}{99}
\bibitem{Ritsch2002Cooling}
P. Domokos and H. Ritsch, 
Collective Cooling and Self-Organization of Atoms in a Cavity, 
Phys. Rev. Lett. \textbf{89}, 253003 (2002).


\bibitem{ritsch2013cold}
H. Ritsch, P. Domokos, F. Brennecke, and T. Esslinger,
Cold atoms in cavity-generated dynamical optical potentials,
Rev. Mod. Phys. \textbf{85}, 553 (2013).

\bibitem{mivehvar2021cavity}
F. Mivehvar, F. Piazza, T. Donner, and H. Ritsch,
Cavity QED with quantum gases: new paradigms in many-body physics,
Adv. Phys. \textbf{70}, 1 (2021).





\bibitem{baumann2010dicke}
K. Baumann, C. Guerlin, F. Brennecke, and T. Esslinger,
Dicke quantum phase transition with a superfluid gas in an optical cavity,
Nature \textbf{464}, 1301 (2010).

\bibitem{WuHaiBing2021FermionSuperradiance}
X. Zhang, Y. Chen, Z. Wu, J. Wang, J. Fan, S. Deng, and H. Wu, Observation of a superradiant quantum phase transition in an intracavity degenerate Fermi gas,
Science \textbf{373}, 1359 (2021).

\bibitem{Nagy2010SR}
D. Nagy, G. K\'{o}nya, G. Szirmai, and P. Domokos, 
Dicke-Model Phase Transition in the Quantum Motion of a Bose-Einstein Condensate in an Optical Cavity,
Phys. Rev. Lett. \textbf{104}, 130401 (2010).

\bibitem{Zhai2014SR-Fermi}
Y. Chen, Z. Yu, and H. Zhai,
Superradiance of Degenerate Fermi Gases in a Cavity, 
Phys. Rev. Lett. \textbf{112}, 143004 (2014).



\bibitem{Keeling2014SR-Fermi}
J. Keeling, M. J. Bhaseen, and B. D. Simons, 
Fermionic Superradiance in a Transversely Pumped Optical Cavity,
Phys. Rev. Lett. \textbf{112}, 143002 (2014).

\bibitem{Piazza2014SR-Fermi}
F. Piazza and P. Strack,
Umklapp Superradiance with a Collisionless Quantum Degenerate Fermi Gas,
Phys. Rev. Lett. \textbf{112}, 143003 (2014).


\bibitem{Lv2024-SR}
G.-L. Zhu, C.-S. Hu, H. Wang, W. Qin, X.-Y. L\"u, and F. Nori,
Nonreciprocal Superradiant Phase Transitions and Multicriticality in a Cavity QED System,
Phys. Rev. Lett. \textbf{132}, 193602 (2024).

\bibitem{Zhang2021-SR-RabiTriangle}
Y.-Y. Zhang, Z.-X. Hu, L. Fu, H.-G. Luo, H. Pu, and X.-F. Zhang,
Quantum Phases in a Quantum Rabi Triangle,
Phys. Rev. Lett. \textbf{127}, 063602 (2021).

\bibitem{Zhang2022-SR-RabiRing}
D. Fallas Padilla, H. Pu, G.-J. Cheng, and Y.-Y. Zhang,
Understanding the Quantum Rabi Ring Using Analogies to Quantum Magnetism,
Phys. Rev. Lett. \textbf{129}, 183602 (2022).



\bibitem{Hemmerich2015SR-SF-MI}
J. Klinder, H. Keßler, M. R. Bakhtiari, M. Thorwart, and A.
Hemmerich, 
Observation of a Superradiant Mott Insulator in the Dicke-Hubbard Model, 
Phys. Rev. Lett. \textbf{115}, 230403 (2015).

\bibitem{Esslinger2016MI-Cavity}
R. Landig, L. Hruby, N. Dogra, M. Landini, R. Mottl, T. Donner and T. Esslinger,
Quantum phases from competing short- and long-range interactions in an optical lattice,
Nature \textbf{532}, 476 (2016).




\bibitem{Esslinger2012Roton}
R. Mottl, F. Brennecke, K. Baumann, R. Landig, T. Donner, and T. Esslinger,
Roton-type mode softening in a quantum gas with cavity-mediated long-range interactions, 
Science \textbf{336}, 1570 (2012).

\bibitem{Lev2010SuperSolid}
S. Gopalakrishnan, B. L. Lev, and P. M. Goldbart,
Atom-light crystallization of Bose-Einstein condensates in multimode cavities: Nonequilibrium classical and quantum phase transitions, emergent lattices, supersolidity, and frustration, 
Phys. Rev. A \textbf{82}, 043612 (2010).



\bibitem{Hemmerich2022CTC}
P. Kongkhambut, J. Skulte, L. Mathey, J. G. Cosme, A. Hemmerich, H. Ke{\ss}ler,
Observation of a continuous time crystal,
Science \textbf{377}, 670 (2022).


\bibitem{Keeling2010LC}
J. Keeling, M. J. Bhaseen, and B. D. Simons,
Collective Dynamics of Bose-Einstein Condensates in Optical Cavities,
Phys. Rev. Lett. 105, 043001 (2010)

\bibitem{Risch2015LC}
F. Piazza and H. Ritsch, 
Self-Ordered Limit Cycles, Chaos, and Phase Slippage with a Superfluid inside an Optical Resonator,
Phys. Rev. Lett. \textbf{115}, 163601 (2015).




\bibitem{dreon2021self}
D. Dreon, A. Baumg\"artner, X. Li, S. Hertlein, T. Esslinger, and T. Donner,
Self-oscillating pump in a topological dissipative atom-cavity system,
Nature \textbf{608}, 494 (2022).






























\bibitem{laughlin1981quantized}
R. B. Laughlin,
Quantized Hall conductivity in two dimensions,
Phys. Rev. B \textbf{23}, 5632 (1981).



\bibitem{rice1982elementary}
M. J. Rice and E. J. Mele,
Elementary excitations of a linearly conjugated diatomic polymer,
Phys. Rev. Lett. \textbf{49}, 1455 (1982).



\bibitem{Mancini2015synthetic}
M. Mancini, G. Pagano, G. Cappellini, L. Livi, M. Rider, J. Catani, C. Sias, P. Zoller, M. Inguscio, M. Dalmonte, and L. Fallani,
Observation of chiral edge states with neutral fermions in synthetic Hall ribbons, 
Science \textbf{349}, 1510 (2015).

\bibitem{Spielman2015synthetic}
B. K. Stuhl, H.-I. Lu, L. M. Aycock, D. Genkina, and I. B. Spielman, 
Visualizing edge states with an atomic Bose gas in the quantum Hall regime,
Science \textbf{349}, 1514 (2015).

\bibitem{Boada2012synthetic}
O. Boada, A. Celi, J. I. Latorre, and M. Lewenstein,
Quantum simulation of an extra dimension,
Phys. Rev. Lett. \textbf{108}, 133001 (2012).


\bibitem{Zilberberg2013synthetic}
Y. E. Kraus, Z. Ringel, and O. Zilberberg,
Four-dimensional quantum Hall effect in a two-dimensional quasicrystal, 
Phys. Rev. Lett. \textbf{111}, 226401 (2013).


\bibitem{celi2014synthetic}
A. Celi, P. Massignan, J. Ruseckas, N. Goldman, I. B. Spielman, G. Juzeli\=unas, and M. Lewenstein,
Synthetic gauge fields in synthetic dimensions,
Phys. Rev. Lett. \textbf{112}, 043001 (2014).


\bibitem{Goldman2015synthetic}
H. M. Price, O. Zilberberg, T. Ozawa, I. Carusotto, and N. Goldman,
Four-dimensional quantum Hall effect with ultracold atoms, Phys. Rev. Lett. \textbf{115}, 195303 (2015).

\bibitem{Lustig2019photonic}
E. Lustig, S. Weimann, Y. Plotnik, Y. Lumer, M. A. Bandres, A. Szameit and M. Segev,
Topological photonics in synthetic dimensions,
Nature \textbf{567}, 356 (2019).

\bibitem{Dutt2020photonic}
A. Dutt, Q. Lin, L. Yuan, M. Minkov, M. Xiao, and S. Fan, 
A single photonic cavity with two independent physical synthetic dimensions,
Science \textbf{367}, 59 (2020).


\bibitem{Fan2016photonic}
Q. Lin, M. Xiao, L. Yuan, and S. Fan, 
Photonic Weyl point in a two-dimensional resonator lattice with a synthetic frequency dimension, 
Nat. Commun. \textbf{7}, 13731 (2016).


\bibitem{Fan2018photonic}
L. Yuan, Q. Lin, M. Xiao, and S. Fan, 
Synthetic dimension in photonics, 
Optica \textbf{5}, 1396 (2018).


\bibitem{Ozawa2019Rev}
T. Ozawa, H. M. Price, A. Amo, N. Goldman, M. Hafezi, L. Lu, M. C. Rechtsman, D. Schuster, J. Simon, O. Zilberberg, I. Carusotto, 
Topological photonics,
Rev. Mod. Phys. \textbf{91}, 015006 (2019).



\bibitem{xiao2010berry}
D. Xiao, M.-C. Chang, and Q. Niu,
Berry phase effects on electronic properties,
Rev. Mod. Phys. \textbf{82}, 1959 (2010).


\bibitem{thouless1982quantized}
D. J. Thouless, M. Kohmoto, M. P. Nightingale, and M. den Nijs,
Quantized Hall conductance in a two-dimensional periodic potential,
Phys. Rev. Lett. \textbf{49}, 405 (1982).

\bibitem{Halperin2017photonic}
I. Martin, G. Refael, and B. Halperin,
Topological frequency conversion in strongly driven quantum systems,
Phys. Rev. X \textbf{7}, 041008 (2017).



\bibitem{Esslinger2021imbalanced}
X. Li, D. Dreon, P. Zupancic, A. Baumgärtner, A. Morales, W. Zheng, N. R. Cooper, T. Donner, and T. Esslinger,
First order phase transition between two centro-symmetric superradiant crystals,
Phys. Rev. Research \textbf{3}, L012024 (2021).


\bibitem{ZW2023imbalanced}
X. Nie and W. Zheng,
Nonequilibrium phases of a Fermi gas inside a cavity with imbalanced pumping,
Phys. Rev. A \textbf{108}, 043312 (2023).

\bibitem{SM}
See Supplemental Material for (a) The atomic current operator in real space direction; (b) Mapping the Hamiltonian from atom-cavity Hilbert space to synthetic two dimensional lattice and the current operator in synthetic direction; (c) Effective Hamiltonian in the zero-photon sector at large detuning regime. 






  










\end{thebibliography}
\end{document}


\title{Supplemental Material for ``Synthetic Berry curvature in atom-cavity systems''}

\author{Zheng Tang}
\thanks{These authors contributed equally to this work.}
\affiliation{State Key Laboratory of Quantum Optics Technologies and Devices, Institute of Theoretical Physics, Shanxi University, Taiyuan 030006, China}
\author{Rui-Lin Zhang}
\thanks{These authors contributed equally to this work.}
\affiliation{Hefei National Research Center for Physical Sciences at the Microscale and School of Physical Sciences, 
University of Science and Technology of China, Hefei 230026, China}
\author{Xiaotian Nie}
\affiliation{Intelligent Quantum Inception Co., Ltd., Haidian, Beijing, 100083, China}
\affiliation{iFLYTEK Research, Hefei, 230088, China}
\author{Li Chen}
\email{lchen@sxu.edu.cn}
\affiliation{State Key Laboratory of Quantum Optics Technologies and Devices, Institute of Theoretical Physics, Shanxi University, Taiyuan 030006, China}
\author{Wei Zheng}
\email{zw8796@ustc.edu.cn}
\affiliation{Hefei National Research Center for Physical Sciences at the Microscale and School of Physical Sciences, 
University of Science and Technology of China,
Hefei 230026, China}
\affiliation{CAS Center for Excellence in Quantum Information and Quantum Physics,
University of Science and Technology of China, Hefei 230026, China}
\affiliation{Hefei National Laboratory, 
University of Science and Technology of China, Hefei 230088, China}

\maketitle

In this Supplemental Material, we derive the atomic current operator in real space direction. Then we map the Hamiltonian from atom-cavity Hilbert space to synthetic two dimensional lattice. Based on the Hamiltonian in synthetic lattice, we obtain the current operator in synthetic direction, and demonstrate it is the increasing rate of the photon number. Finally, we derive the effective Hamiltonian in the zero-photon sector at large detuning regime, and identify its current-like term. 

\section{The current operator}
The Hamiltonian of the main text is
\begin{align}
\hat H={}&\Delta\hat a^\dagger\hat a+
\Bigl(m+\delta m\hat P\Bigr)\hat D
-\Bigl(h-\delta h\hat Q\Bigr)\hat K_0
-\Bigl(h+\delta h\hat Q\Bigr)\hat K_1,
\label{eq:SM_H}
\end{align}
where $\hat{Q}=\bigl(\hat a+\hat a^\dagger \bigr)/2$, $\hat{P}= \bigl(\hat a-\hat a^\dagger \bigr)/2\ii$, and 
\begin{align}
\hat D   &=\sum_i\left(\hat c_{iA}^\dagger\hat c_{iA}-\hat c_{iB}^\dagger\hat c_{iB} \right), \\
\hat K_0 &=\sum_i\left(\hat c_{iB}^\dagger\hat c_{iA}+\hc \right), \\
\hat K_1 &=\sum_i\left(\hat c_{i+1,A}^\dagger\hat c_{iB}+\hc \right).
\label{eq:SM_defs}
\end{align}
Here $i=1,\ldots,L$ labels unit cells, $A,B$ are sublattices, and $i+L\equiv i$ implements periodic real-space boundaries. The annihilation and creation operators of cavity mode are $\hat{a}$ and $\hat{a}^{\dagger}$. We define $\hat K_\pm=\hat K_0\pm\hat K_1$, and write the Hamiltonian into 
\begin{align}
\hat H&=\Delta\hat a^\dagger\hat a+\hat H_{\rm at} +\hat a^\dagger\hat V+\hat a\hat V^\dagger, 
\label{eq:SM_decomp}
\end{align}
where 
\begin{align}
\hat {H}_{\mathrm{at}} &= m\hat D - h\hat K_+, \\
\hat {V} &=\frac{1}{2}\left(\delta h\hat K_-+\ii\delta m\hat D \right). 
\end{align}
The last equality is exact: $\hat a^\dagger\hat V+\hat a\hat V^\dagger=\delta h\hat Q\hat K_-+\delta m\hat P\hat D$.

Following the bond-phase convention, assign the same dimensionless Peierls phase $\phi$ to both rightward bonds, $A_i\to B_i$ and $B_i\to A_{i+1}$:
\begin{align}
\hat K_0(\phi)&=\sum_i \left(e^{\ii\phi}\hat c_{iB}^\dagger\hat c_{iA}+\hc \right),\\
\hat K_1(\phi)&=\sum_i \left(e^{\ii\phi}\hat c_{i+1,A}^\dagger\hat c_{iB}+\hc \right),
\label{eq:SM_peierlsx}
\end{align}
The phase around the ring is $\Phi=2L\phi$; $\phi$ itself is a phase per bond. Equal phases correspond to equal bond lengths in a uniform vector potential. The Hamiltonian becomes
\begin{equation}
\hat H(\phi)=\Delta\hat n_{\rm ph}+\Big(m+\delta m\hat P \Big)\hat D
 -\Big( h-\delta h\hat Q \Big)\hat K_0(\phi)
 -\Big( h+\delta h\hat Q \Big)\hat K_1(\phi).
\end{equation}
Here $\hat{n}_{\mathrm{ph}} = \hat{a}^{\dagger}\hat{a}$ is the photon number. The variation of the Hamiltonian with respect to the Peierls phase yields the atomic current operator
\begin{equation}
\hat J_x=-\frac{1}{L}\left.\frac{\partial\hat H(\phi)}{\partial\phi}\right|_{\phi=0}.
\label{eq:SM_Jxdef}
\end{equation}
Since
\begin{align}
\left. \frac{\partial\hat K_0(\phi)}{\partial\phi}\right|_0 &=\ii\sum_i \left(\hat c_{iB}^\dagger\hat c_{iA}-\hc \right),\\
\left. \frac{\partial\hat K_1(\phi)}{\partial\phi}\right|_0 &=\ii\sum_i \left(\hat c_{i+1,A}^\dagger\hat c_{iB}-\hc \right),
\end{align}
we obtain $\hat{J}_{x} = \hat{J}_{0} + \hat{J}_{1}$, where 
\begin{equation}
\begin{aligned}
\hat J_{0} &=\frac{\ii}{L}\left(h-\delta h\hat Q \right)
\sum_i\left(\hat c_{iB}^\dagger\hat c_{iA}-\hc \right)\\
\hat J_{1} &= \frac{\ii}{L}\left(h+\delta h\hat Q \right)
\sum_i \left(\hat c_{i+1,A}^\dagger\hat c_{iB}-\hc \right).
\end{aligned}
\label{eq:SM_Jx}
\end{equation}
$\hat J_0$ and $\hat J_1$ are the cell-averaged intracell and intercell currents, respectively. 

\section{Map to the synthetic dimension}
\subsection{Single-particle hopping Hamiltonian}
We first consider the single-atom problem. Let $\ket{i,\tau}$ denote a single-atom orbit located at unit cell $i$ and sublattice $\tau$. Therefore the atomic operators, $\hat{D}$, $\hat{K}_{0}$, and $\hat{K}_{1}$, can be expressed in single-atom case as 
\begin{align}
\hat D^{(1)}&=\sum_i \Bigl( \ket{i,A}\bra{i,A}-\ket{i,B}\bra{i,B} \Bigr),\\
\hat K_0^{(1)}&=\sum_i \Bigl( \ket{i,B}\bra{i,A}+\hc \Bigr), \\
\hat K_1^{(1)}&=\sum_i \Bigl( \ket{i+1,A}\bra{i,B}+\hc \Bigr).
\end{align}
On another hand, the annihilation and creation operators of cavity mode can be rewritten into the Fock-ladder form
\begin{align}
\hat {a}               &= \sum_{n=0}^\infty \sqrt{n} \ket {n-1} \bra{n}, \\
\hat {a}^{\dagger}     &= \sum_{n=0}^\infty \sqrt{n} \ket{n} \bra {n-1}, \\
\hat {n}_{\mathrm{ph}} & = \sum_{n=0}^\infty n \ket{n} \bra {n},
\label{eq:SM_ladder}
\end{align}
Then we define the single-particle orbit in synthetic two dimensional lattice as
\begin{equation}
\ket{n;i,\tau}=\ket {n} \otimes\ket{i,\tau}, \qquad n=0,1,2,\ldots.
\label{eq:SM_basis}
\end{equation}
Here $(i,\tau)$ denote the coordinate in physical direction, while $n$ denote the coordinate in synthetic direction. Therefore, the single-atom Hamiltonian can be rewritten into
\begin{align}
\hat H^{(1)}={}&\Delta\hat n_{\rm ph}\otimes\hat I_{\rm at}
+\left(m\hat I_{\rm ph}+\delta m\frac{\hat a-\hat a^\dagger}{2\ii}\right)\otimes\hat D^{(1)} \nonumber\\
&-\left(h\hat I_{\rm ph}-\delta h\frac{\hat a+\hat a^\dagger}{2}\right)\otimes\hat K_0^{(1)}
-\left(h\hat I_{\rm ph}+\delta h\frac{\hat a+\hat a^\dagger}{2}\right)\otimes\hat K_1^{(1)}.
\end{align}
where $\hat{I}_{\mathrm{ph}}= \sum_{n=0}^{\infty} \ket{n}\bra{n}$ and $\hat{I}_{\mathrm{at}} = \sum_{i,\tau}\ket{i,\tau}\bra{i,\tau}$ are identity operator in Hilbert space of cavity and atoms respectively. Inserting the Fock expansions gives explicitly
\begin{align}
\hat H^{(1)}={}&\sum_{n=0}^{\infty}\ket n\bra n\otimes
  \left(\Delta n\hat I_{\rm at}+m\hat D^{(1)}-h\hat K_+^{(1)} \right) \nonumber\\
&-\frac{\ii\delta m}{2}\sum_{n=0}^{\infty}\sqrt n
  \Bigl(\ket{n-1}\bra n-\ket n\bra{n-1} \Bigr) \otimes\hat D^{(1)} \nonumber\\
&+\frac{\delta h}{2}\sum_{n=0}^{\infty}\sqrt n
  \Bigl(\ket{n-1}\bra n+\ket n\bra{n-1} \Bigr)\otimes
  \left(\hat K_0^{(1)}-\hat K_1^{(1)} \right).
\label{eq:SM_tensor}
\end{align}
Each factor is converted to a product-basis hopping by
$(\ket{n'}\bra n)\otimes(\ket{i',\tau'}\bra {i,\tau}) =\ket{n'; i', \tau'}\bra{n; i, \tau}$. This yields
\begin{align}
\hat H^{(1)}={}&\sum_{n=0}^\infty\sum_i\Bigl[(\Delta n+m)\ket{n;i,A}\bra{n;i,A}
+(\Delta n-m)\ket{n;i,B}\bra{n;i,B}\Bigr]\nonumber\\
&-h\sum_{n=0}^\infty\sum_i\Bigl(\ket{n;i,B}\bra{n;i,A} +\ket{n;i+1,A}\bra{n;i,B}+\hc\Bigr)\nonumber\\
&-\frac{\ii\delta m}{2}\sum_{n=1}^\infty\sum_i\sqrt n
\Bigl(\ket{n-1;i,A}\bra{n;i,A}
-\ket{n-1;i,B}\bra{n;i,B}-\hc\Bigr)\nonumber\\
&+\frac{\delta h}{2}\sum_{n=1}^\infty\sum_i\sqrt n\Bigl(
\ket{n-1;i,B}\bra{n;i,A}-\ket{n-1;i+1,A}\bra{n;i,B}\nonumber\\
&\hspace{30mm}+\ket{n;i,B}\bra{n-1;i,A}
-\ket{n;i+1,A}\bra{n-1;i,B}+\hc\Bigr).
\label{eq:SM_singleH}
\end{align}
The first line is the layer-dependent on-site energy. The second line gives real-space hopping within one photon layer. The $\delta m$ term connects the same atomic orbital on adjacent layers with opposite imaginary amplitudes on the two sublattices. The $\delta h$ term connects opposite sublattices while changing photon number; its intracell and intercell amplitudes have opposite signs. Both upward and downward diagonal paths are present. Their $\sqrt n$ amplitude is the Bose-enhancement factor. Sums that change layers start at $n=1$ so no unphysical state $\ket{-1}$ occurs. These are the note's $n\geq0$ sums with their vanishing $n=0$ term removed.

For $\Delta>0$, the on-site potential increases along $n$ and the associated force points toward smaller $n$: $F_n=-\Delta$ in these units. Ignoring Bose enhancement makes neighboring-layer hopping uniform, but translation symmetry additionally requires $\Delta=0$ and removal of the vacuum boundary, for example in an auxiliary periodic synthetic lattice.

For the case with many atoms, one can replace $\ket{n;i,\tau}\bra{n';j,\tau'}$ by $\hat{\tilde c}_{n;i\tau}^\dagger\hat{\tilde c}_{n';j\tau'}$ gives
\begin{align}
\hat H_{\rm syn}=&\sum_{n=0}^{\infty}\sum_i
\Bigl[(\Delta n+m)\hat{\tilde c}_{n;iA}^\dagger\hat{\tilde c}_{n;iA}
+(\Delta n-m)\hat{\tilde c}_{n;iB}^\dagger\hat{\tilde c}_{n;iB} \Bigr]\nonumber\\
&-h\sum_{n=0}^{\infty}\sum_i
\Bigl(\hat{\tilde c}_{n;iB}^\dagger\hat{\tilde c}_{n;iA}
+\hat{\tilde c}_{n;i+1,A}^\dagger\hat{\tilde c}_{n;iB}+\hc \Bigr)\nonumber\\
&-\frac{\ii\delta m}{2}\sum_{n=1}^{\infty}\sum_i\sqrt n
\Bigl(\hat{\tilde c}_{n-1;iA}^\dagger\hat{\tilde c}_{n;iA}
-\hat{\tilde c}_{n-1;iB}^\dagger\hat{\tilde c}_{n;iB}-\hc \Bigr)\nonumber\\
&+\frac{\delta h}{2}\sum_{n=1}^{\infty}\sum_i\sqrt n
\Bigl[\hat{\tilde c}_{n-1;iB}^\dagger\hat{\tilde c}_{n;iA}
-\hat{\tilde c}_{n-1;i+1,A}^\dagger\hat{\tilde c}_{n;iB}\nonumber\\
&\hspace{30mm}+\hat{\tilde c}_{n;iB}^\dagger\hat{\tilde c}_{n-1;iA}
-\hat{\tilde c}_{n;i+1,A}^\dagger\hat{\tilde c}_{n-1;iB}+\hc\Bigr].
\label{eq:SM_second}
\end{align}
However, since all atoms share the same photon number, a physical constrain must be imposed,
\begin{equation}
\hat{\tilde{c}}^{\dagger}_{n,i\tau}\hat{\tilde{c}}^{\dagger}_{n',i'\tau'}=0,
\qquad n \ne n',
\end{equation}
That indicate the composite fermions cannot occupy different photon coordinates simultaneously. Extending this quadratic Hamiltonian to unrestricted many-particle occupations would describe a different system.

\subsection{Synthetic Peierls phase}
To describe the current along the synthetic direction, we impose the Peierls phase $\theta$ in the photon ladder. Attach $e^{-\ii\theta}$ to each lowering transition and $e^{+\ii\theta}$ to each raising transition. Let $\hat H_{\rm diag}$ denote the first two lines of Eq.~\eqref{eq:SM_second}. Then
\begin{align}
\hat H(\theta)={}&\hat H_{\rm diag}
-\frac{\ii\delta m}{2}\sum_{n=1}^{\infty}\sum_i\sqrt n
\Bigl[e^{-\ii\theta}\hat{\tilde c}_{n-1;iA}^\dagger\hat{\tilde c}_{n;iA}
-e^{-\ii\theta}\hat{\tilde c}_{n-1;iB}^\dagger\hat{\tilde c}_{n;iB}-\hc\Bigr]\nonumber\\
&+\frac{\delta h}{2}\sum_{n=1}^{\infty}\sum_i\sqrt n
\Bigl[e^{-\ii\theta}\hat{\tilde c}_{n-1;iB}^\dagger\hat{\tilde c}_{n;iA}
-e^{-\ii\theta}\hat{\tilde c}_{n-1;i+1,A}^\dagger\hat{\tilde c}_{n;iB}\nonumber\\
&\hspace{34mm}+e^{+\ii\theta}\hat{\tilde c}_{n;iB}^\dagger\hat{\tilde c}_{n-1;iA}
-e^{+\ii\theta}\hat{\tilde c}_{n;i+1,A}^\dagger\hat{\tilde c}_{n-1;iB}+\hc\Bigr].
\label{eq:SM_Htheta}
\end{align}
In particular, the last forward term raises $n$ and therefore carries $e^{+\ii\theta}$, irrespective of its real-space direction. One can calculate the current operator along the synthetic direction by variation of the Hamiltonian with respect to the corresponding Peierls phase $\theta$ 
\begin{equation}
\hat J_n=-\left.\frac{\partial\hat H(\theta)}{\partial\theta}\right|_{\theta=0}.
\label{eq:SM_Jnq}
\end{equation}
Therefore, we obtain 
\begin{align}
\hat J_{n} &=\frac{\delta m}{2}\sum_{n=1}^{\infty}\sum_i\sqrt n
\Bigl(\hat{\tilde c}_{n-1;iA}^\dagger\hat{\tilde c}_{n;iA}
-\hat{\tilde c}_{n-1;iB}^\dagger\hat{\tilde c}_{n;iB} + \hc \Bigr)\nonumber\\
&-\ii\frac{\delta h}{2}\sum_{n=1}^{\infty}\sum_i\sqrt n
\Bigl[\hat{\tilde c}_{n-1;iB}^\dagger\hat{\tilde c}_{n;iA}
-\hat{\tilde c}_{n-1;i+1,A}^\dagger\hat{\tilde c}_{n;iB}\nonumber\\
&\hspace{30mm}+\hat{\tilde c}_{n;iB}^\dagger\hat{\tilde c}_{n-1;iA}
-\hat{\tilde c}_{n;i+1,A}^\dagger\hat{\tilde c}_{n-1;iB} - \hc\Bigr].
\label{eq:SM_Htheta}
\end{align}
Mapping back from synthetic two dimensional lattice to the atom-cavity Hilbert space, we have 
\begin{equation}
\begin{aligned}
\hat J_n & =\delta m\frac{\hat a+\hat a^\dagger}{2}\hat D
-\delta h\frac{\hat a-\hat a^\dagger}{2\ii}\hat{K}_{-}  \\
& =\delta m\hat Q\hat D-\delta h\hat P\hat K_-.
\end{aligned}
\label{eq:SM_Jn}
\end{equation}
This result has an independent exact check in the original atom-cavity Hilbert space. Since
$[\hat n_{\rm ph},\hat a^\dagger]=\hat a^\dagger$ and
$[\hat n_{\rm ph},\hat a]=-\hat a$, Eq.~\eqref{eq:SM_decomp} implies
\begin{align}
\frac{\dd\hat n_{\rm ph}}{\dd t}
& =-\ii\bigl[\hat n_{\rm ph},\hat H\bigr] \\
& =-\ii\bigl(\hat a^\dagger\hat V-\hat a\hat V^\dagger \bigr)\nonumber\\
& =\delta m\hat Q\hat D-\delta h\hat P\hat K_- \\ & =\hat J_n.
\label{eq:SM_Jndef}
\end{align}
Equivalently, $\hat H(\theta)=\Delta\hat n_{\rm ph}+\hat H_{\rm at}
+e^{+\ii\theta}\hat a^\dagger\hat V+e^{-\ii\theta}\hat a\hat V^\dagger$
has the same derivative. Positive current increases photon number. The detuning is diagonal in $n$ and therefore does not contribute directly to the commutator. A stationary eigenstate with finite photon expectation has $\langle\hat J_n\rangle=0$, while under driving the integrated current gives the change of photon number.

\section{Effective Hamiltonian in the zero-photon sector}
We now return to the original atom--cavity Hilbert space. For a low-energy, zero-photon description we take $\Delta>0$. A negative detuning makes the untruncated photon ladder unbounded below and does not support this ground-state interpretation.

We assume weak couplings compared with detuning $\delta m, \delta h \ll\Delta$, such one can find the effective Hamiltonian projected to the zero-photon sector. Define projectors by
\begin{equation}
\hat\Pi=\ket0\bra0\otimes\hat I_{\rm at}, \\
\hat\Theta = \hat{I} -\hat\Pi.
\label{eq:SM_PT}
\end{equation}
For $\ket\Psi=\ket\psi+\ket\chi$, with $\ket\psi=\hat\Pi\ket\Psi$ and $\ket\chi=\hat\Theta\ket\Psi$, projection of $\hat H\ket\Psi=E\ket\Psi$ gives
\begin{align}
\hat\Pi\hat H\hat\Pi\ket\psi+\hat\Pi\hat H\hat\Theta\ket\chi&=E\ket\psi,\\
\hat\Theta\hat H\hat\Pi\ket\psi+\hat\Theta\hat H\hat\Theta\ket\chi&=E\ket\chi.
\end{align}
When the complementary resolvent exists,
\begin{equation}
\ket\chi = \big(E-\hat\Theta\hat H\hat\Theta \big)^{-1}\hat\Theta\hat H\hat\Pi\ket\psi.
\end{equation}
Substitution into the first equation gives the exact, energy-dependent Feshbach expression, acting within $\hat\Pi$,
\begin{equation}
\hat H_{\rm eff}(E)=\hat\Pi\hat H\hat\Pi+
\hat\Pi\hat H\hat\Theta\frac{1}{E-\hat\Theta\hat H\hat\Theta}
\hat\Theta\hat H\hat\Pi.
\label{eq:SM_feshbach}
\end{equation}
The blocks following from Eq.~\eqref{eq:SM_decomp} are
\begin{align}
\hat\Pi\hat H\hat\Pi&=\ket0\bra0\otimes\hat H_{\rm at},\\
\hat\Theta\hat H\hat\Pi&=\ket1\bra0\otimes\hat V,\\
\hat\Pi\hat H\hat\Theta&=\ket0\bra1\otimes\hat V^\dagger.
\label{eq:SM_blocks}
\end{align}
Only the one-photon block of the resolvent enters to second order. To the lowest order the inverse on the entire complementary space can be replaced by a scalar, and one obtain an energy-independent effective Hamiltonian as
\begin{align}
\hat H_{\rm eff} &=\hat H_{\rm at}+\delta\hat H_{\rm eff}^{(2)},\\
\delta\hat H_{\rm eff}^{(2)}&=-\frac{1}{\Delta}\hat V^\dagger\hat V.
\label{eq:SM_Heff}
\end{align}
The latter includes excursions beyond the one-photon intermediate state. The sequence is $0\xrightarrow{\hat V}1\xrightarrow{\hat V^\dagger}0$, which fixes the ordering $\hat V^\dagger\hat V$.
Using Eq.~\eqref{eq:SM_decomp}, we have 
\begin{equation}
\delta\hat H_{\rm eff}^{(2)}=
-\frac{\delta h^2}{4\Delta}\hat K_-^2
-\frac{\delta m^2}{4\Delta}\hat D^2
+\frac{\ii\delta h\delta m}{4\Delta}\bigl[\hat D,\hat K_- \bigr].
\label{eq:SM_Heffexpand}
\end{equation}
The final term is Hermitian because the commutator is anti-Hermitian. The first two terms arise from using the same coupling channel twice; the last combines one bond-coupling event with one density-coupling event. The two possible orderings differ because the atomic operators do not commute.

To make its current content explicit, we calculate
\begin{align}
\bigl[\hat D,\hat K_0 \bigr]&=-2\sum_i \left(\hat c_{iB}^\dagger\hat c_{iA}-\hc \right),\\
\bigl[\hat D,\hat K_1 \bigr]&=2\sum_i \left(\hat c_{i+1,A}^\dagger\hat c_{iB}-\hc \right),
\end{align}
and therefore
\begin{align}
\bigl[\hat D,\hat K_- \bigr]=-2\Biggl[&\sum_i \left(\hat c_{iB}^\dagger\hat c_{iA}-\hc \right)
+\sum_i \left(\hat c_{i+1,A}^\dagger\hat c_{iB}-\hc \right)\Biggr].
\label{eq:SM_comm}
\end{align}
At $n=0$, $\hat\Pi\hat Q\hat\Pi=0$, so Eq.~\eqref{eq:SM_Jx} becomes
\begin{align}
\hat\Pi\hat J_x\hat\Pi=\frac{\ii h}{L}\hat\Pi\Biggl[&\sum_i \left(\hat c_{iB}^\dagger\hat c_{iA}-\hc \right)
+\sum_i \left(\hat c_{i+1,A}^\dagger\hat c_{iB}-\hc \right)\Biggr]\hat\Pi.
\end{align}
Combining this equation with Eq.~\eqref{eq:SM_comm} yields
\begin{equation}
\frac{\ii\delta h\delta m}{4\Delta}\bigl[\hat D,\hat K_- \bigr] \propto -\hat\Pi\hat J_x\hat\Pi, 
\label{eq:SM_currentrelation}
\end{equation}
Both sides act in the same zero-photon subspace. That indicate a finite current carrying state would lower the energy of the effective Hamiltonian.